\documentclass[amsmath,amssymb,floatfix,lengthcheck,preprintnumbers,prl,showpacs,superscriptaddress,twocolumn]{revtex4}
\usepackage{graphicx}
\usepackage{slashed}
\usepackage{xcolor}
\newcommand{\beq}{\begin{equation}}
\newcommand{\eeq}{\end{equation}}
\newcommand{\beqs}{\begin{eqnarray}}
\newcommand{\eeqs}{\end{eqnarray}}

\definecolor{red}{rgb}{1.0, 0, 0}

\begin{document}

\title{Lattice calculation of composite dark matter form factors}

\author{T.~Appelquist}
\affiliation{Department of Physics, Sloane Laboratory, Yale University,
             New Haven, Connecticut 06520, USA}
\author{R.~C.~Brower}
\affiliation{Department of Physics, Boston University,
	Boston, Massachusetts 02215, USA}
\author{M.~I.~Buchoff}
\affiliation{Lawrence Livermore National Laboratory, Livermore, California 94550, USA}
\author{M.~Cheng}
\affiliation{Department of Physics, University of Washington, Box 351560, Seattle, WA 98195, USA}
\author{S.~D.~Cohen}
\affiliation{Department of Physics, University of Washington, Box 351560, Seattle, WA 98195, USA}
\author{G.~T.~Fleming}
\affiliation{Department of Physics, Sloane Laboratory, Yale University,
             New Haven, Connecticut 06520, USA}
\author{J.~Kiskis}
\affiliation{Department of Physics, University of California,
	Davis, California 95616, USA}
\author{M.~F.~Lin}
\affiliation{Department of Physics, Sloane Laboratory, Yale University,
             New Haven, Connecticut 06520, USA}
\author{E.~T.~Neil}
\affiliation{Theoretical Physics Department, Fermi National Accelerator Laboratory, 
        Batavia, IL 60510, USA}
\author{J.~C.~Osborn}
\affiliation{Argonne Leadership Computing Facility,
	Argonne, Illinois 60439, USA}
\author{C.~Rebbi}
\affiliation{Department of Physics, Boston University,
	Boston, Massachusetts 02215, USA}
\author{D.~Schaich}
\affiliation{Department of Physics,
        University of Colorado, Boulder, CO 80309, USA}
\author{C.~Schroeder}
\affiliation{Lawrence Livermore National Laboratory, Livermore, California 94550, USA}
\author{S.~Syritsyn}
\affiliation{%
Lawrence Berkeley National Laboratory, Berkeley, CA 94720, USA}
\author{G.~Voronov}
\affiliation{Department of Physics, Sloane Laboratory, Yale University,
             New Haven, Connecticut 06520, USA}
\author{P.~Vranas}
\affiliation{Lawrence Livermore National Laboratory, Livermore, California 94550, USA}
\author{J.~Wasem}
\affiliation{Lawrence Livermore National Laboratory, Livermore, California 94550, USA}
\collaboration{Lattice Strong Dynamics (LSD) Collaboration}
\noaffiliation

\begin{abstract}
Composite dark matter candidates, which can arise from new strongly-coupled sectors, are
well-motivated and phenomenologically interesting, particularly in the context of asymmetric
generation of the relic density.  In this work, we employ lattice calculations to study the
electromagnetic form factors of electroweak-neutral dark-matter baryons for a three-color,
QCD-like theory with $N_f = 2$ and $6$ degenerate fermions in the  fundamental representation.
We calculate the (connected) charge radius and anomalous magnetic moment, both of which can play
a significant role for direct detection of composite dark matter. We find minimal $N_f$
dependence in these quantities. We generate mass-dependent cross-sections for dark
matter-nucleon interactions and use them in conjunction with experimental results from XENON100,
excluding dark matter candidates of this type with masses below $10\text{ TeV}$.
\end{abstract}

\preprint{FERMILAB-PUB-13-014-T, LLNL-JRNL-608695, NT-LBL-13-002, UCB-NPAT-13-002}

\maketitle

\paragraph{\textbf{Introduction}}

Experimental bounds on the interaction of the dark matter with Standard-Model (SM) particles
have strengthened by many orders of magnitude in recent years.  In particular, dark-matter
particles cannot have SM-strength couplings to electroweak gauge bosons, based on
direct-detection constraints \cite{Goodman:1984dc, Kurylov:2003ra}.  At the same time, there is
a strong motivation for the dark matter to couple to the SM in some way for the purpose of relic
density generation, either as a thermal relic via the so-called ``WIMP miracle" (see
\cite{Hooper:2009zm} for a recent review) or through an asymmetric scenario which may be related
to the creation of baryon asymmetry
\cite{Nussinov:1985xr,Barr:1990ca,Kaplan:1991ah,Gudnason:2006ug,Gudnason:2006yj,Nardi:2008ix,Kaplan:2009ag,Shelton:2010ta}. Construction of
dark matter models thus requires a careful balance between the presence and absence of
dark-sector interactions with the SM.

Composite dark matter models provide a simple mechanism for attaining this balance, one which
can lead to interesting and unique phenomenology.  By hypothesizing a new, confining gauge force
in the dark sector, an electroweak-neutral composite dark matter candidate can be constructed as
a bound state of electroweak-charged constituents.  In this way, electroweak interactions can be
active in the early Universe for the generation of relic density, but only neutral bound states
survive to the present day. Electroweak coupling to the constituents is still possible, leading
to form-factor suppressed interactions with the neutral composites. They can be roughly
estimated from QCD analogs, but in general can be determined quantitatively only by lattice
calculations.

In this paper, we consider an underlying $SU(3)$ gauge theory with 
fermions in the fundamental representation, but focus on fermions not 
associated with electroweak breaking.
We use $SU(3)$ because much is known about it from lattice QCD and 
because we have already generated lattice vacuum states of $SU(3)$
with 2 and 6 fundamental flavors on large 
lattices~\cite{Appelquist:2009ka, Appelquist:2010xv}.
We take the fermions to be mass-degenerate $SU(2)_L$ singlets such that $Q=Y$.
We consider a two-fermion theory
($N_f = 2$) with $Q_u = 2/3$ and $Q_d = -1/3$, as well as a six-fermion
theory ($N_f = 6$) with three such pairs of fermions. In either case,
the lightest baryon is expected to be electrically neutral, and will
therefore also have vanishing weak charge. The dominant contribution to
its interaction with ordinary nuclei will be due to single photon
exchange, which can be parameterized primarily in terms of its magnetic
moment and charge radius. 
In these initial lattice calculations we consider only quark-line connected 
contributions to the charge radius and magnetic moment.
We compute the electromagnetic form factors of
this particle to extract these quantities, describe their dependence on
$N_f$, and discuss consequences for direct detection.

One could also modify or enlarge the fermion content of the $SU(3)$ gauge theory to include
$SU(2)_L$-doublet fermions.  This would be a necessary modification in order to consider
composite dark matter arising in a theory of dynamical electroweak symmetry breaking~\cite{
Nussinov:1985xr,Chivukula:1989qb,Barr:1990ca,Kaplan:1991ah,Bagnasco:1993st,
Gudnason:2006ug,Gudnason:2006yj,Foadi:2008qv,Nardi:2008ix,Frandsen:2009mi,Banks:2010eh, 
Kribs:2009fy,Fok:2011yc}.
Careful model building is then required to ensure that the lightest baryon 
is net electroweak neutral.
We do not discuss this possibility here.

\paragraph{\textbf{Model setup}}

For the theory with $SU(2)_L$-singlet fermions carrying charges $Q_u = 2/3$ and $Q_d = -1/3$,
with $N_f = 2$ or $6$, the analogue of the neutron ($N \sim udd$) will be the dark matter
candidate, with mass $M_B$ and carrying no net electroweak charge.  It is stabilized by
conservation of dark baryon number.  The other charged baryons are expected to be heavier due to
electromagnetic mass corrections of order $\Delta M \sim \alpha M_B / 4\pi$.  We include a
fermion mass  $m_f$, essential for lattice calculation purposes, and examine dependence on $m_f$
for a range $m_f \ll M_B$.

Our dark sector also contains $N_f^2 - 1$ pseudo-Nambu-Goldstone-boson (PNGB) states.  We assume
that these states are unstable, decaying to Standard-Model particles with a sufficient rate that
their presence does not influence the cosmological history of the Universe.

As our focus is on direct-detection signatures, we do not consider the dark matter generation in
detail here.  The confinement scale $\Lambda$, or equivalently the dark matter mass $M_B$, is a
free parameter in our construction.

\paragraph{\textbf{Electromagnetic Form Factors}}
Since the neutral baryon in the $SU(2)$-singlet theory is the dark matter
candidate of interest \footnote{
 Depending on other symmetry considerations,
 the pseudo-Goldstone bosons could also play the role of dark matter
 \cite{Belyaev:2010kp,Bai:2010qg,Frigerio:2012uc,Buckley:2012ky}.},
the baryon mass $M_B$ (degenerate in the absence of other
interactions) is the dark matter mass. This mass and all other dimensionful
quantities are expressed in lattice units here.

The quantities of central interest here are the Dirac and the Pauli electromagnetic form factors
of a neutral dark-matter baryon $| N(p) \rangle$. For the $N_f = 2$ case, they can be expressed
in terms of matrix elements of the vector currents of individual quarks as follows:
\begin{equation}
\label{eqn:emff_def}
\begin{aligned}
& \langle N(p^\prime) | \overline{\psi} \gamma^\mu \psi | N(p) \rangle\\
&\quad = \overline{U}(p^\prime)\Bigg[F_1^\psi(Q^2)\gamma^\mu
   + F_2^\psi(Q^2)\frac{i\sigma^{\mu\nu}q_\nu}{2M_B}\Bigg]U(p)\,,
\end{aligned}
\end{equation}
where $\psi = u, d$ are quark fields, $U,\,\overline{U}$ are on-shell baryon spinors,
$q=p^\prime - p$, and $Q^2 = -q^2 > 0$ is the momentum transfer. 
In the forward limit $Q^2=0$, the Dirac form factors are equal 
to the numbers of the valence quarks: $F_1^u(0)=1$ and $F_1^d(0)=2$.

From these one constructs the isovector and isoscalar form factors\footnote{
  Sign conventions of the isovector and isoscalar form factors are chosen to agree 
  with nuclear physics notations; since we use the neutron ($udd$) as our 
  in- and out-states,   we have to swap $u\longleftrightarrow d$ in the r.h.s of 
  Eq.~\ref{eqn:iso_formfac}.}:
\begin{equation}
\label{eqn:iso_formfac}
\begin{aligned}
F^v_{1,2}(Q^2) &= F^d_{1,2}(Q^2)-F^u_{1,2}(Q^2)\,,\\
F^s_{1,2}(Q^2) &= F^d_{1,2}(Q^2)+F^u_{1,2}(Q^2)\,.
\end{aligned}
\end{equation}
Both of these quantities can be extracted from lattice calculations, but the
isoscalar contribution contains expensive disconnected lattice quark contractions, 
which cancel in the isovector case, and as a result, isovector form factors are far more
tractable.  While we ultimately will
calculate the disconnected pieces of the isoscalar form factor as well, this
work will focus on only the connected contributions.

For the $N_f = 6$ case, with three pairs of $u (Q=2/3)$ and $d (Q = -1/3)$ fermions,
we take the $| N(p) \rangle$ state to be composed of fermions from only one pair.
Since we omit disconnected lattice quark contractions in our calculation,
it is only the currents $\overline{\psi}\gamma^\mu\psi$ composed of the fermion fields 
from the same pair that contribute to the computed electromagnetic form factors.
Therefore, in our calculation the other two pairs play a role in only 
the strong dynamics of the $SU(3)$ gauge theory.

The full electromagnetic form factors of the neutral dark baryon\footnote{
  We denote the observables associated with the neutral dark baryon in our calculations
  with the subscript ``neut'' to avoid confusion with the QCD neutron, for which we reserve
  the subscript ``n''.} 
are given by
\begin{equation}
\begin{aligned}
F_{1,2;\text{neut}}(Q^2) &= Q_u F^u_{1,2}(Q^2) + Q_d F^d_{1,2}(Q^2) \\
&= \frac{1}{6} F^s_{1,2}(Q^2) - \frac{1}{2} F^v_{1,2}(Q^2)\,;
\end{aligned}
\end{equation}
since $F_1^s(0)=3$ and $F_1^v(0)=1$, the total charge $F_{1;\text{neut}}(0) = 0$.  
For soft single-photon exchange scattering, only the forward ($Q^2\to0$) behavior 
of the electromagnetic form factors is relevant.
Since the electric charge $F_{1;\text{neut}}(0)$ is zero, 
only the magnetic moment $\mu_\text{neut}=\kappa_\text{neut}$ 
and the Dirac radius $\langle r_{1;\text{neut}}^2\rangle$ contribute 
to the scattering amplitude to the lowest order in $Q^2$:
\begin{equation}
\label{eqn:kappa_def}
\begin{aligned}
F_{1;\text{neut}}(Q^2) &= -\frac16 Q^2\langle r_{1;\text{neut}}^2\rangle + {\mathcal O}(Q^4)\,, \\
F_{2;\text{neut}}(Q^2) &= \kappa_\text{neut} + {\mathcal O}(Q^2)\,,
\end{aligned}
\end{equation}
The Dirac charge radius $\langle r_{1;\text{neut}}^2\rangle$ determines 
the slope of the form factor in the $Q^2\to0$ limit:
\begin{equation}
\label{eqn:r1neutral_def}
\langle r_{1;\text{neut}}^2\rangle \overset{def}{=} 
  -6\frac{dF_{1;\text{neut}}(Q^2)}{dQ^2}\Big|_{Q^2=0}\,.
\end{equation}
The definition of the radius (\ref{eqn:r1neutral_def}) is motivated by the algebraic identity
\begin{equation}
\label{eqn:r1neutral_equiv}
\int d^3 r \, r^2 \, \rho(r)
  \equiv -6\frac{dF_1(Q^2)}{dQ^2}\Big|_{Q^2=0}\,,
\end{equation}
where $\rho(r)$ is the ``charge density'',
\begin{equation}
\int d^3r \, e^{i\vec q\vec r} \, \rho(r) = F_1(Q^2)\,,
\quad Q^2\underset{non-rel.}{\approx}\vec q^2\,,
\end{equation}
which has physical meaning if and only if the spatial extent of this distribution 
is much larger than the Compton wave length of the composite particle, 
$\langle r^2 \rangle  \gg M_B^{-2}$.
Since the total charge, $\int d^3r\,\rho(r) \equiv F_1(0)$, is zero, the charge density
must have alternating sign (or be exactly zero), and the integral in
Eq.~(\ref{eqn:r1neutral_equiv}) can be either positive or negative.

For the following, we also need to define the \emph{mean squared charge radius} 
$\langle r_E^2\rangle$, or the ``radius'' of the charge form factor $G_E(Q^2)$,
\begin{equation}
G_E (Q^2) = F_1(Q^2) - \frac{Q^2}{4M_B^2} F_2(Q^2)\,.
\end{equation}
Similar to Eq.~(\ref{eqn:r1neutral_def}), the charge radius of the neutral baryon is equal to
\begin{equation}
\langle r_{E;\text{neut}}^2\rangle 
  \overset{def}{=} -6\frac{dG_{E;\text{neut}}(Q^2)}{dQ^2}\Big|_{Q^2=0}
  = \langle r_{1;\text{neut}}^2\rangle + \frac{3\kappa_\text{neut}}{2M_B^2}\,,
\end{equation}
differing from the Dirac radius by only the relativistic correction $\sim M_B^{-2}$ 
(the Foldy term).
This correction is important if the size of the particle is comparable to its Compton wave length,
which is the case for the neutron and the proton in QCD.

The (anomalous) magnetic moment of the neutral baryon is related to the isovector and isoscalar
moments as
\begin{equation}
\kappa_\text{neut} = \frac16\kappa_s - \frac12\kappa_v\,.
\end{equation}
The isovector and isoscalar Dirac form factors are not zero in the forward limit.
Their radii are defined to be independent of their overall normalization,
\begin{equation}
\label{eqn:ff_lowQ2}
F_1^{v,s}(Q^2) = F_1^{v,s}(0)\big[1 -\frac16 Q^2 \langle r_1^2 \rangle^{v,s} + O(Q^4)\big]\,.
\end{equation}
The radii of the neutral baryon are related to the isovector and isoscalar radii as follows:
\begin{equation}
\begin{aligned}
 \langle r_{1;\text{neut}}^2\rangle
   &= \frac12\langle r_1^2\rangle^s - \frac12\langle r_1^2\rangle^v\,,\\
 \langle r_{E;\text{neut}}^2\rangle
   &= \frac12\langle r_E^2\rangle^s - \frac12\langle r_E^2\rangle^v\,.
\end{aligned}
\end{equation}

\paragraph{\textbf{Simulation Details}}
Lattice calculations are performed using $32^3 \times 64$ domain-wall lattices with the Iwasaki
improved gauge action and a domain-wall height of $m_0
= 1.8$.  The length of the fifth dimension is fixed at $L_s = 16$.  By using domain-wall fermions, the calculation preserves exact flavor symmetry, and
chiral-breaking lattice spacing artifacts are suppressed.  The
calculation is performed for $N_f = 2$ at $\beta = 2.70$ and $N_f = 6$ at $\beta = 2.10$. The
beta values are tuned to match the confinement scale of both theories relative to the lattice
spacing, including $M_B$ as we shall see below.  
For both $N_f=2$ and $N_f = 6$, five separate mass points are analyzed 
with $m_f = 0.010, 0.015, 0.020, 0.025, 0.030$.
The pion masses (in units of the nucleon mass) are $0.41\le m_\pi/M_B\le0.52$ 
and $0.44\le m_\pi/M_B\le0.52$ for $N_f=2$ and $N_f=6$, respectively.
Further details and other results from these ensembles are given 
in~\cite{Appelquist:2009ka,Appelquist:2010xv,Appelquist:2012sm}.

\paragraph{\textbf{Calculation and Fitting}}
The parameters of interest are extracted from two sets of correlation
functions: two-point correlation functions given by
\begin{equation}
\label{eq:baryon_2_point}
  C_{NN}(\tau,\mathbf{p})=\sum_\mathbf{x} e^{-i\mathbf{p}\cdot \mathbf{x}}
    \langle N(\mathbf{x},\tau) \bar N(0)\rangle,
\end{equation}
and three-point correlation functions
\begin{equation}
\label{eq:baryon_3_point}
\begin{aligned}
C_{N\mathcal{O}N}(\tau,T, \mathbf{p},\mathbf{p^\prime})
  = & \sum_{\mathbf{x},\mathbf{y}}
    e^{-i\mathbf{p^\prime}\cdot\mathbf{x}+i(\mathbf{p^\prime}-\mathbf{p})\cdot\mathbf{y}}
    \times \\
    & \times \langle N(\mathbf{x},T)\mathcal{O}(\mathbf{y},\tau) \bar N(0) \rangle\,,
\end{aligned}
\end{equation}
where $\mathcal{O}(\mathbf{y},\tau)$ is the quark vector current density operator.

The long-distance limit of the Euclidean time behavior of these correlation functions is given by
\begin{gather}
C_{NN}(\tau,\mathbf{p})
  \stackrel{\tau \gg \frac1\Delta}{\longrightarrow}
    \frac{Z(\mathbf{p})e^{-E\tau}}{2E}
    \mathrm{Tr}\Big[\Gamma_\text{pol}(i\slashed{p} + M_B) \Big]\,, \\
\begin{aligned}
C_{N\mathcal{O}N}(\tau,T,\mathbf{p},\mathbf{p^\prime})
  &\stackrel{T, \tau \gg \frac1\Delta}{\longrightarrow}
    \frac{\sqrt{Z(\mathbf{p})Z(\mathbf{p^\prime})}e^{-E^\prime(T-\tau)-E\tau}}
          {4EE^\prime } \times \\
  &\times \mathrm{Tr}\Big[\Gamma_\text{pol} (i\slashed{p}^\prime+ M_B) 
          \Gamma^\mu (i\slashed{p} + M_B)\Big]\,,
\end{aligned}
\end{gather}
where $\Gamma_\text{pol}$ is the polarization matrix of the initial 
and final baryon spin states 
corresponding to Eq.~(\ref{eq:baryon_2_point},\ref{eq:baryon_3_point}),
$\Gamma^\mu$ is the fermion vertex function (cf. Eq.(\ref{eqn:emff_def})), 
\begin{equation}
\Gamma^\mu= F_1(Q^2)\gamma^\mu + F_2(Q^2)\frac{\sigma^{\mu\nu} q_\nu}{2M_B}\,,
\end{equation}
and $\Delta$ is the difference in energy between
the ground and the first excited state of the baryon.
More details on the form factor calculation on the lattice can be found in
Ref.~\cite{Syritsyn:2009mx}.

\begin{figure}[t] 
   \centering
   \includegraphics[width=0.5\textwidth]{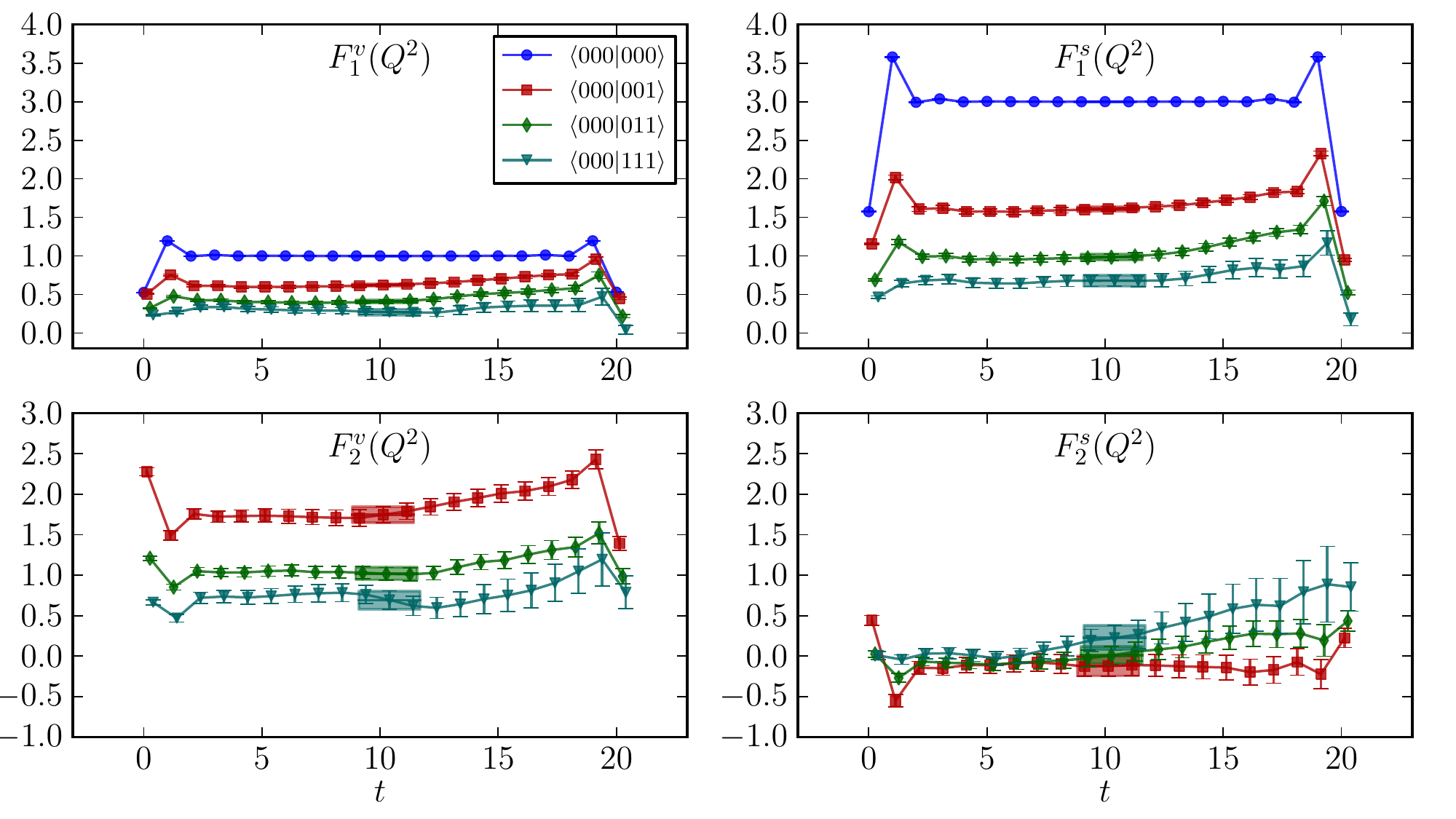}\\
    $(N_f=2)$ \\
   \includegraphics[width=0.5\textwidth]{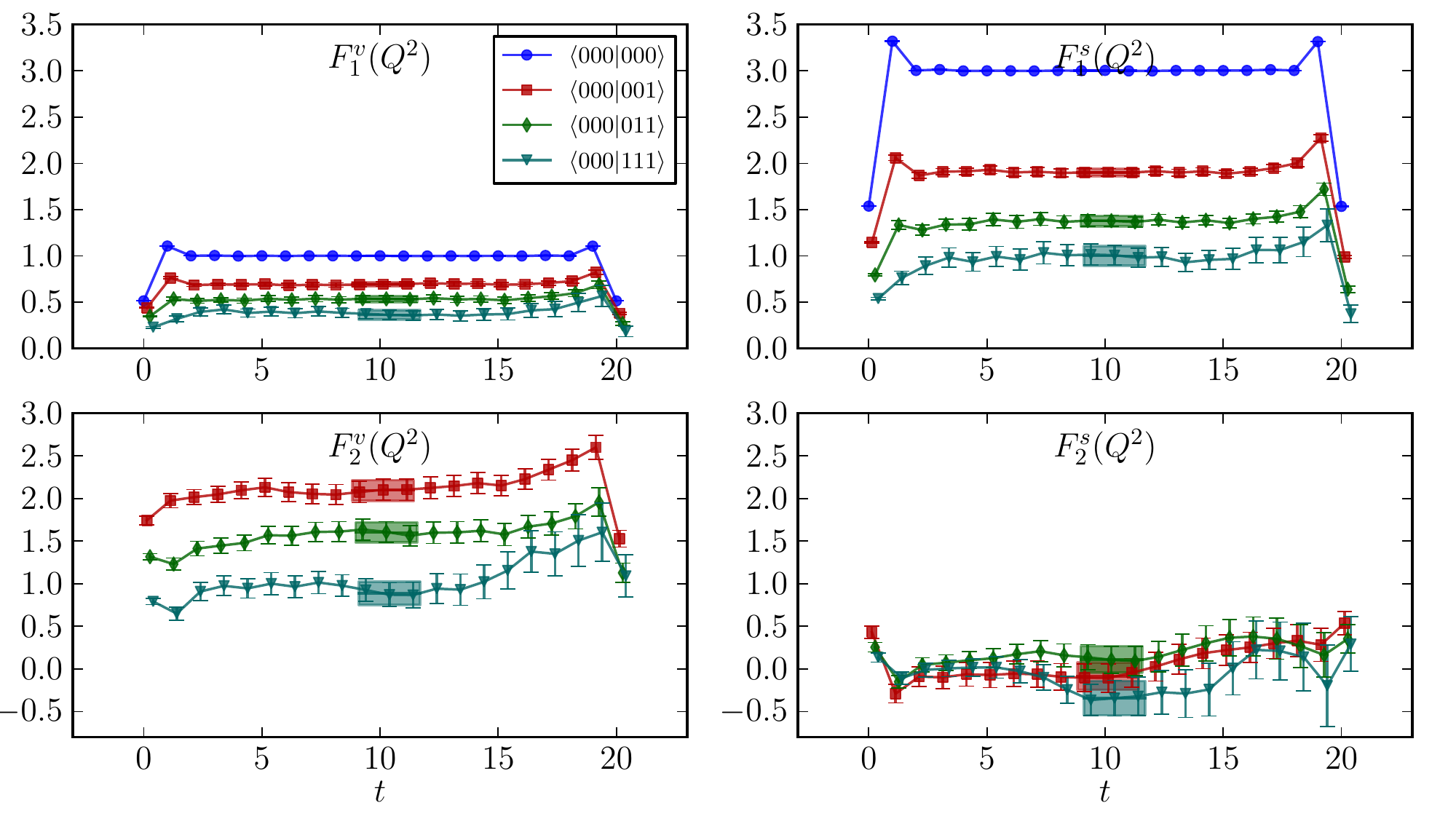}\\
    $(N_f=6)$ \\
   \caption{\label{fig:corr_ratios} 
       Examples of form factor plateaus for $m_f=0.015$ for 2 and 6 flavors.
       Form factor plateaus are shown for all values of $Q^2$ for $F_1$ and 
       for $Q^2>0$ for $F_2$; corresponding lattice initial and final momenta are shown 
       in the legends.}
\end{figure}

We use the standard widely adopted ``ratio'' method in order to extract hadron matrix elements from corresponding two- and three-point functions,
\begin{equation}
\begin{aligned}
R_\mathcal{O}(\tau,T,\mathbf{p},\mathbf{p^\prime})
  &=\frac{C_{N\mathcal{O}N}(\tau,T,\mathbf{p},\mathbf{p^\prime})}
    {\sqrt{C_{NN}(T,\mathbf{p})C_{NN}(T,\mathbf{p^\prime})}}\  \times\\
  &\times\sqrt{\frac{C_{NN}(T-\tau,\mathbf{p})C_{NN}(\tau,\mathbf{p^\prime})}
              {C_{NN}(T-\tau,\mathbf{p^\prime})C_{NN}(\tau,\mathbf{p})}},
\end{aligned}
\end{equation}
where the long Euclidean time behavior yields
\begin{equation}
\label{eqn:corr_ratios}
\begin{gathered}
R_\mathcal{O}(\tau,T,\mathbf{p},\mathbf{p^\prime})
  \stackrel{T, \tau \gg \frac1\Delta}{\longrightarrow}
    \langle N(\mathbf{p^\prime}) | \mathcal{O} | N(\mathbf{p}) \rangle
  \quad\quad\\
  \quad\quad + \mathcal{O}(e^{-\Delta \tau}) + \mathcal{O}(e^{-\Delta (T - \tau)})
    +\mathcal{O}(e^{-\Delta T})
\end{gathered}
\end{equation}
We analyze these ratios for multiple initial and final momentum combinations and vector current components in order to extract form factors $F_1$ and $F_2$. Their values form a reasonable ``plateau''  as functions of $\tau$, the timeslice of the current operator insertion, indicating absence of significant excited-state contaminations (see Fig.~\ref{fig:corr_ratios}).

In general, excited states can cause significant systematic errors in three-point functions and hadron matrix elements~\cite{Green:2011fg}. We compute our form factor values as averages of three central points in the plateaus.

\begin{figure}[t] 
   \centering
   \includegraphics[width=0.5\textwidth]{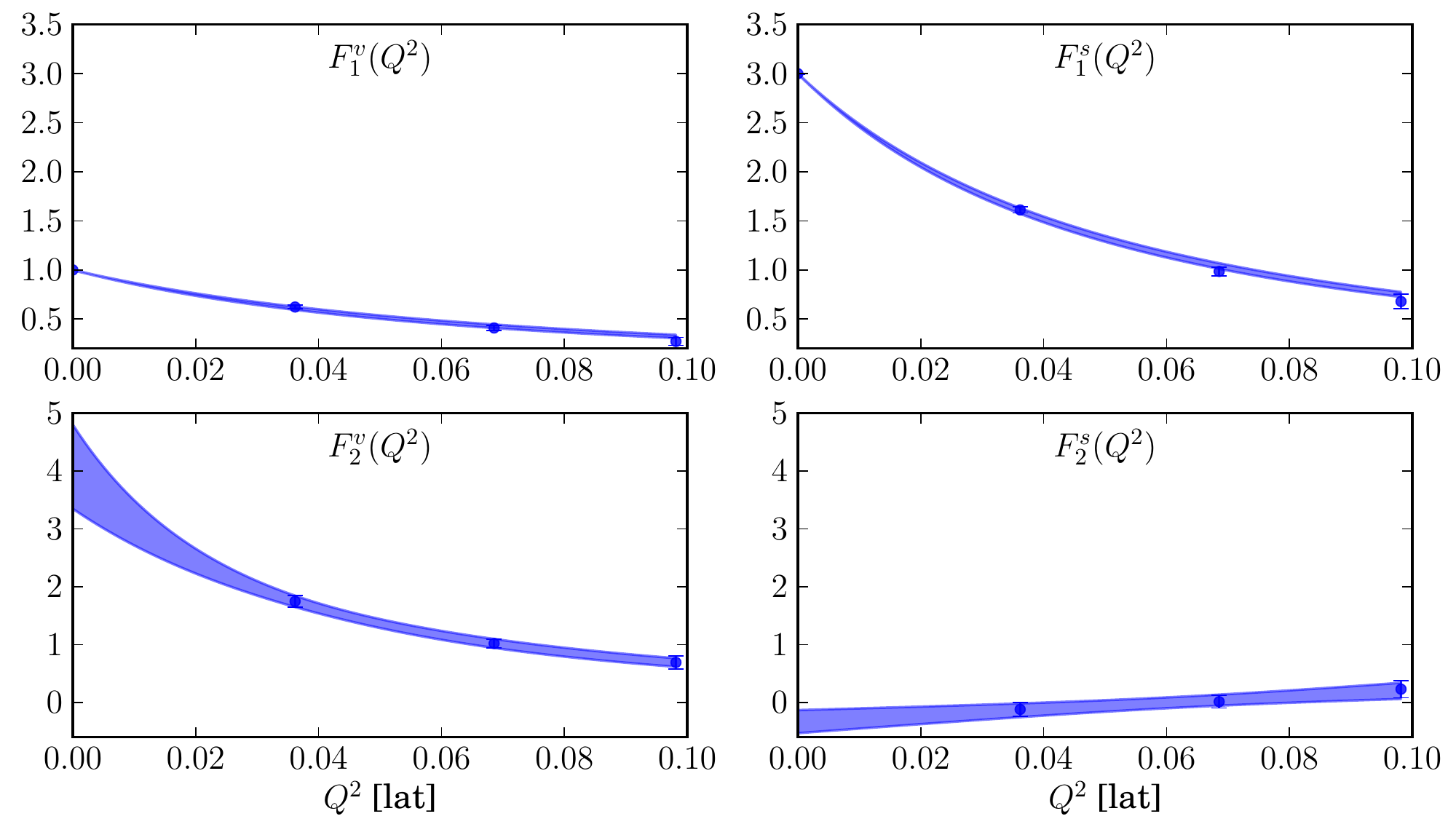}\\
    $(N_f=2)$ \\
   \includegraphics[width=0.5\textwidth]{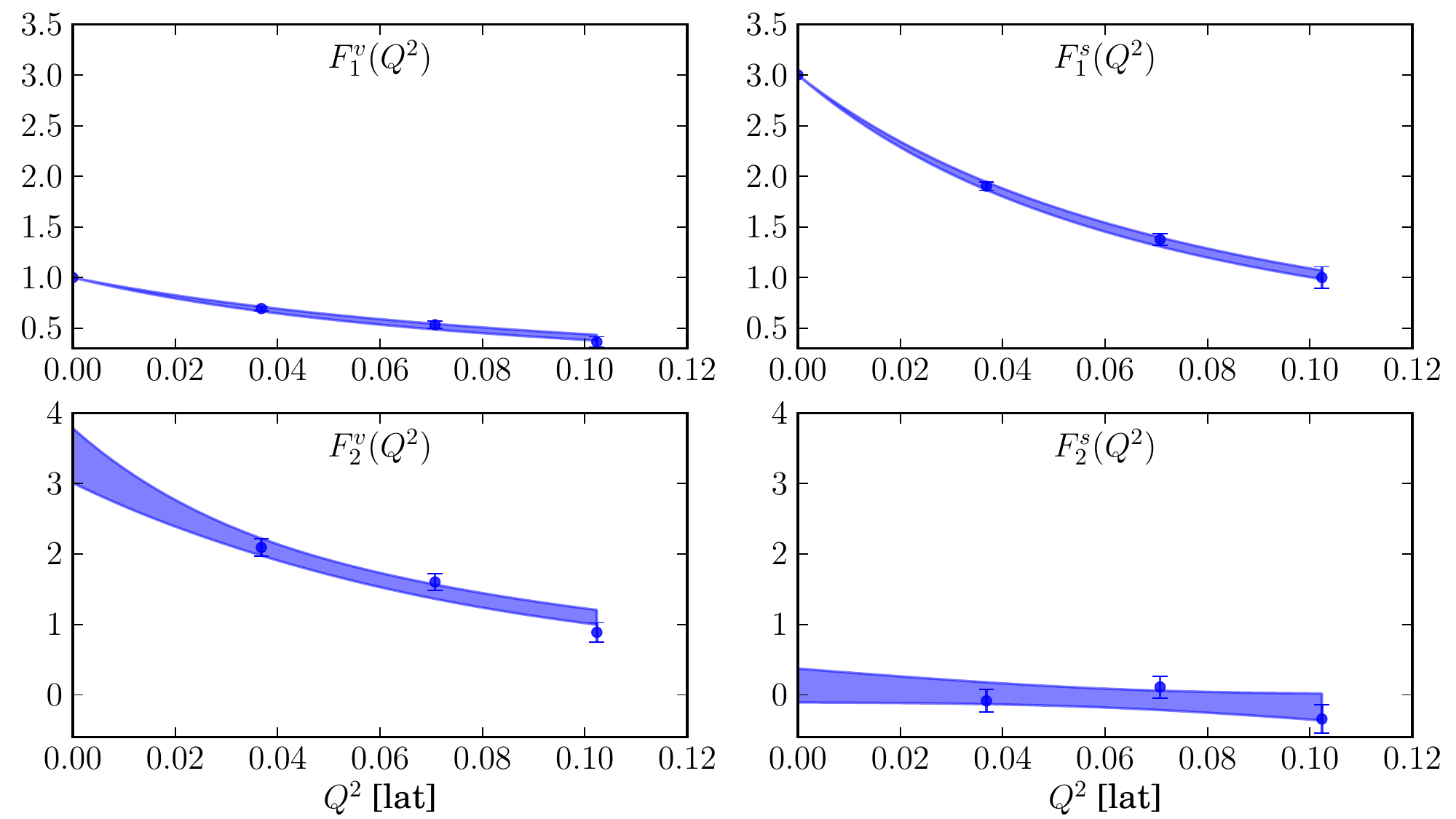}\\
    $(N_f=6)$ \\
   \caption{\label{fig:ff_q2_fits}
       Examples of $Q^2$ fits of Dirac and Pauli form factors $F_{1,2}(Q^2)$ 
       for $m_f=0.015$ for 2 and 6 flavors.
       The bands show the dipole fits for all form factors except $F_2^{s}$, 
       for which we used the linear fit for $Q^2\to0$ extrapolation.}
\end{figure}

The form factors $F_{1,2}(Q^{2})$ are calculated at discrete values of the momentum transfer
$Q^2\approx(\mathbf{p^\prime}-\mathbf{p})^2$ determined by the lattice volume.
We interpolate the Dirac and isovector Pauli form factors using a dipole formula fit 
\begin{equation}
F_{1,2}(Q^2)\sim\frac {A_{1,2}} {(1+B_{1,2}Q^2)^2}
\end{equation}
motivated by nucleon form factor phenomenology.
The isoscalar Pauli form factor turns out to be very close to zero, 
and the dipole form that has definite sign does not necessarily yield a stable fit to the data;
therefore, we use the linear fit $F_i(Q^2) \sim F_{i}(0) + F_i^{\prime}(0) Q^2 $.
Examples of fits are shown on Fig.~\ref{fig:ff_q2_fits}.
We use these fits to interpolate (extrapolate in the case of Pauli form factors) near
the forward limit $Q^2 = 0$ in order to determine $\kappa$ and $\langle r_1^{2} \rangle$.

\vspace{12pt}
\paragraph{\textbf{Lattice Results}}
\paragraph{\textit{Baryon Mass}}
\begin{figure}[t] 
   \centering
   \includegraphics[width=3.3in]{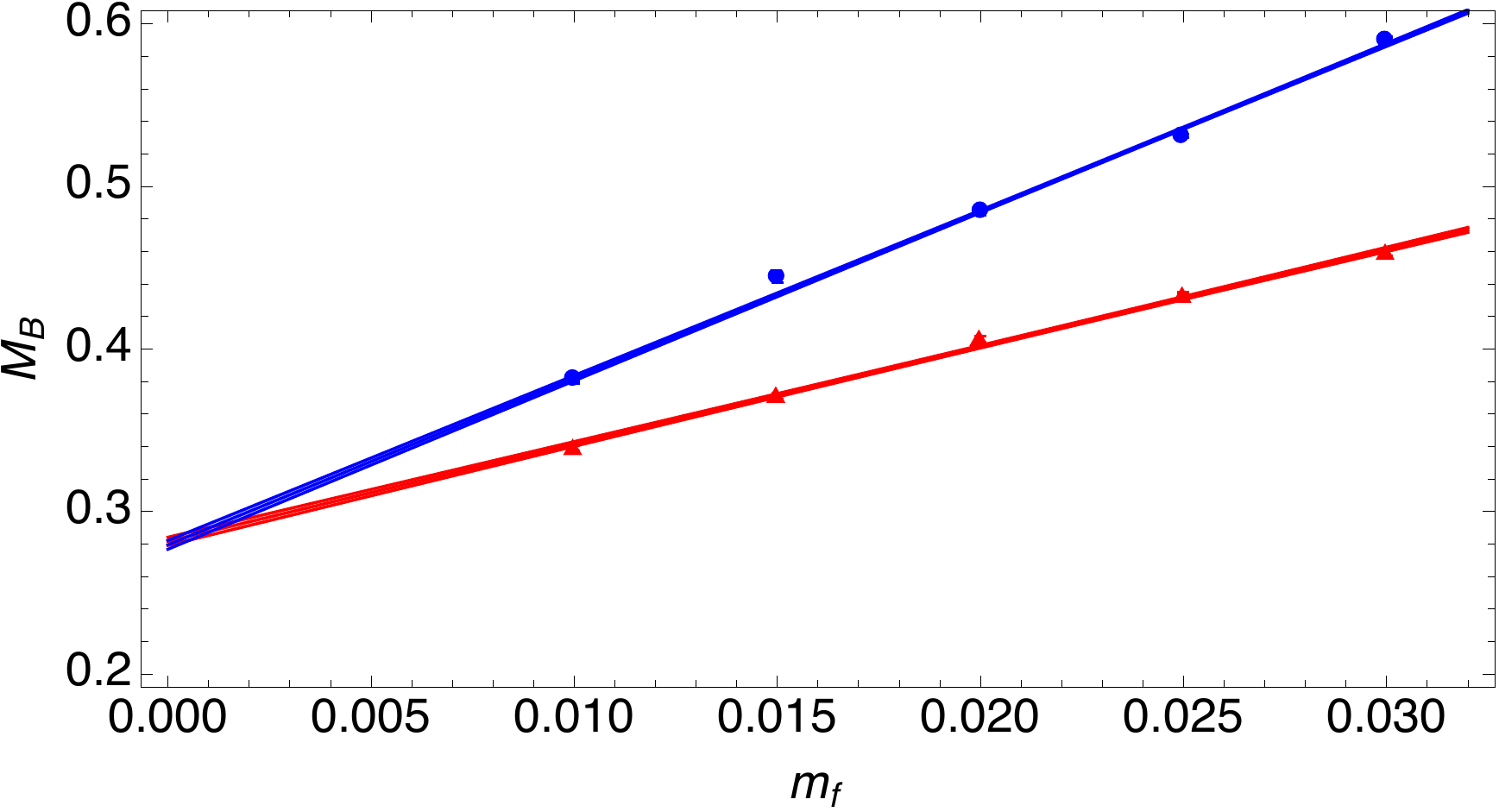}
   \caption{Dark-matter baryon mass (in lattice units) with $N_f = 2$ (red, lower curve) and $N_f = 6$ (blue, upper curve), 
      as a function of the fermion mass $m_f$ (also in lattice units).  
      The two data sets are extrapolated to obtain the chiral-limit baryon
      mass $M_{B_0}$, which is used to set a physical scale independent of $am_f$. 
      With the chosen lattice couplings, $M_{B_0}$ is the same within statistical precision 
      in the $N_f = 2$ and $N_f= 6$ theories.}
   \label{fig:baryon_mass_and_sigma}
\end{figure}
The dark-matter baryon mass is plotted as a function of the fermion mass $m_f$ in
Fig.~\ref{fig:baryon_mass_and_sigma}.  A linear dependence of the baryon mass on $m_f$ can be
seen for both theories, as expected in the calculation regime where the fermion masses are small.
In the absence of additional interactions, a finite value of $m_f$ is required to give mass to
the PNGB's of the theory, but we nevertheless
perform a linear fit in order to extract the chiral-limit baryon mass $M_{B_0}$.  This scale,
which can be taken as a proxy for the confinement scale of the theory, serves as a common
reference scale for the calculation results with $m_f \geq 0$.

\paragraph{\textit{Anomalous magnetic moment}}
\begin{figure}[t] 
  \centering
  \includegraphics[width=3.3in]{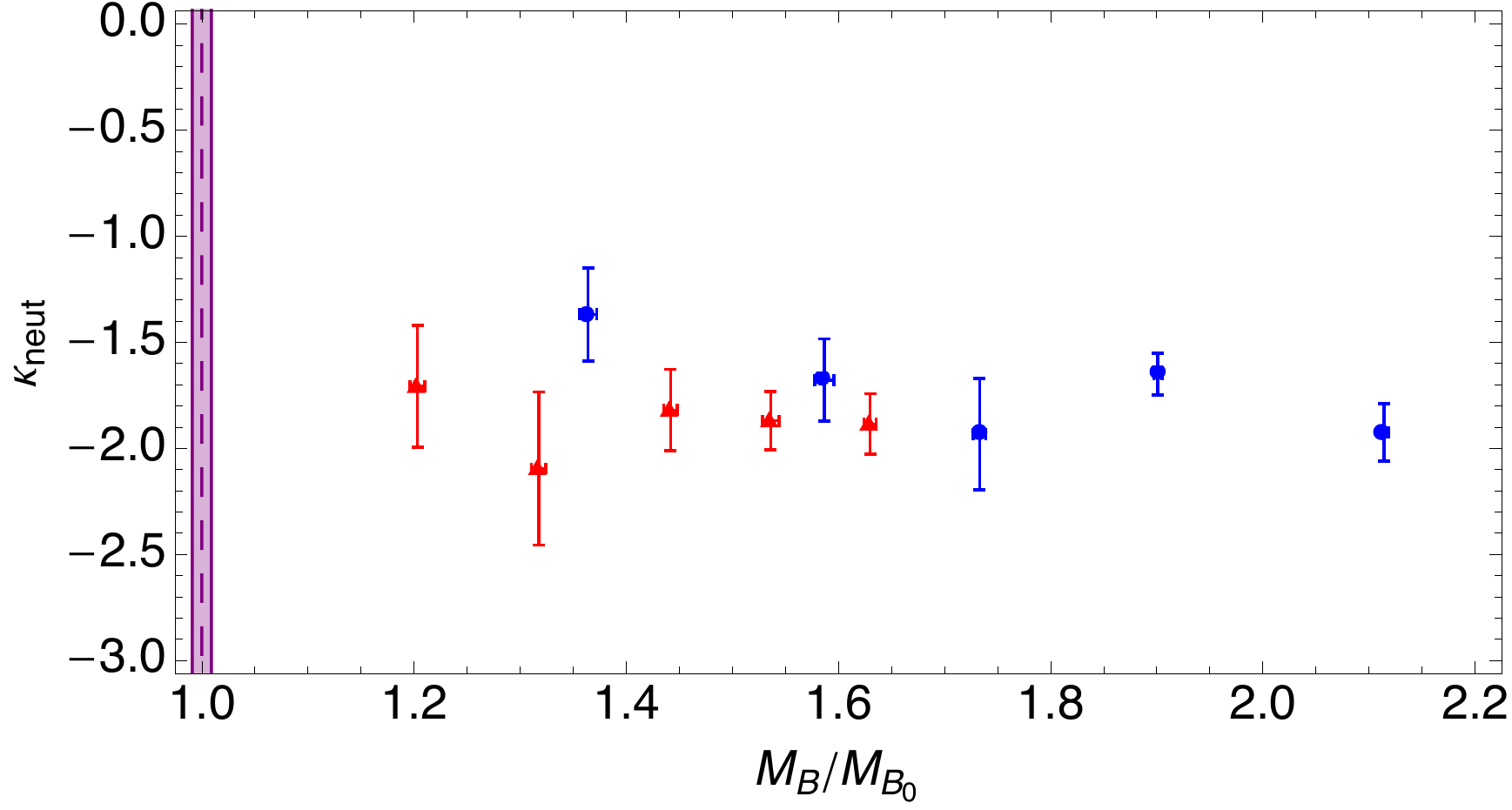}
  \caption{The neutral baryon anomalous magnetic moment 
    for $N_f=2$ (red triangles) and $N_f=6$ (blue circles) theories versus dark-baryon mass.  
    This quantity shows no systematic separation between two and six flavor theories.
    \label{fig:Mag_Moment}}
\end{figure}
The anomalous magnetic moment is the most important for direct detection
experiments. 
It enters at the dimension-5 level in the baryon effective field theory and arises as the
zero-momentum value of the Pauli form factor, $F_2(0)$. 
The isovector Pauli form factor, giving $\kappa_v$, is under most control since all
expensive disconnected contributions cancel due to isospin symmetry. 
The isoscalar channel, which is also necessary to determine $\kappa_\text{neut}$, 
has both connected and disconnected contributions to the three-point correlation function.  
In this initial work, we omit the disconnected
contributions and assume the connected pieces dominate the isoscalar
contribution.

We plot the anomalous magnetic moment $\kappa_\text{neut}$, computed as described above,
versus $M_{B} / M_{B_0}$ in Fig.~\ref{fig:Mag_Moment}.
It shows little dependence on the mass and little dependence on the number of fermions.
The $N_f = 2$ results $\kappa_\text{neut}\approx-(1.71\ldots2.09)$ are consistent with the measured
neutron value $\kappa = -1.91$ \cite{Beringer:1900zz}.
Calculations of nucleon structure with $N_f=2$ Wilson fermions were previously reported
in Ref.\cite{Collins:2011mk}, which found values $\kappa_\text{neut}\approx-(1.30\ldots1.45)$, with the difference coming predominantly from the isovector Pauli form factor; our results for this form factor more closely match the more recent results of \cite{Green:2012ud,Lin:2013bxa}.

\paragraph{\textit{Charge radius}}
\begin{figure}[h] 
  \centering
  \includegraphics[width=3.3in]{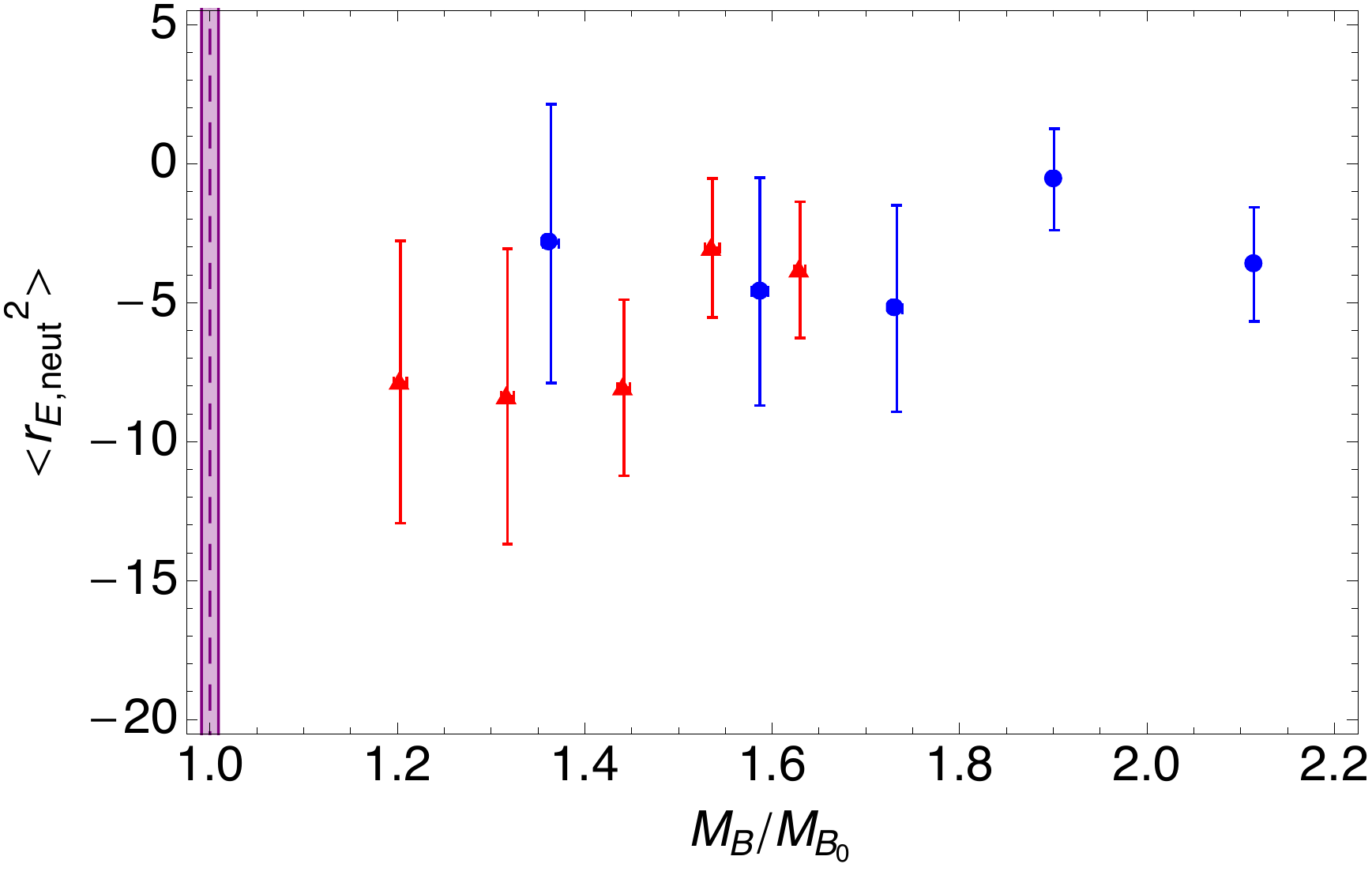}
  \caption{The neutral baryon mean squared charge radius (in lattice units)
    for $N_f = 2$ (red triangles) and $6$ (blue circles), versus dark-baryon mass.  
    Again, no significant systematic difference between the two theories 
    is seen over the range of masses considered.
    \label{fig:Charge_Rad}}
\end{figure}

While the charge radius is expected to lead to a smaller effect on the
spin-independent cross section as compared to the magnetic moment, it could
have a significant effect if its value depends significantly on $N_f$.
It is therefore informative to explore the relative size of the charge radius
contribution to the spin-independent cross section.   As with the magnetic
moment, only the isovector charge radius is absent of disconnected lattice quark contractions,
but we omit them for the isoscalar channel as well.

The results for the mean square charge radius $\langle r_{E;\text{neut}}^2 \rangle$ of
an electroweak-neutral dark-matter baryon are presented in Fig.~\ref{fig:Charge_Rad}.
Note that the results are negative (see discussion after Eq.~(\ref{eqn:r1neutral_equiv})).
As in the case of the anomalous moment, our results show little dependence on $N_f$ and
little dependence on the dark-baryon mass as it varies due to changes in the underlying
fermion mass. 
If the fermion mass is reduced further, bringing $M_B/M_{B_0}$ closer to unity, 
the magnitude $\langle r_{E;\text{neut}}^2 \rangle$ is expected to grow. 
This is because the PNGB mass drops, and the charge radius is quite sensitive 
to the size of the PNGB cloud. 

For $N_f = 2$, this point can be made more precisely by comparison to QCD.
There, the mean squared charge radius of the neutron is also negative, 
$\langle r_{En}^2\rangle=-0.1161(22)\text{ fm}^2$~\cite{Beringer:1900zz}. 
Our $N_f=2$ calculation corresponds to QCD with $M_{B} \approx 1\text{ GeV}$, but with relatively heavy underlying quarks, and thus relatively heavy pions: the pion mass in units of $M_B$ ranges between the lightest $m_\pi/m_B=0.41$ to the heaviest $m_\pi/m_B=0.52$.
In QCD units, our lattice spacing is given by $a \approx 0.055\text{ fm}$, so our result is $\langle r_{E,neut}^2\rangle \approx -(0.009\ldots0.025)\text{ fm}^2$, 
substantially less than the observed result. 
Previous calculations of nucleon structure with $N_f=2$ Wilson fermions~\cite{Collins:2011mk} 
yielded similar values $\langle r_{E,\text{neut}}^2\rangle=-(0.011\ldots0.023)\text{ fm}^2$.
These results, too, employed relatively heavy underlying quarks.
In our case, further studies with smaller fermion mass can shed light on the range of 
direct detection allowed values for the mean square charge radius.

\paragraph{\textbf{Direct detection exclusion plots}}

\begin{figure}[ht!] 
  \centering
  \includegraphics[width=.49\textwidth]{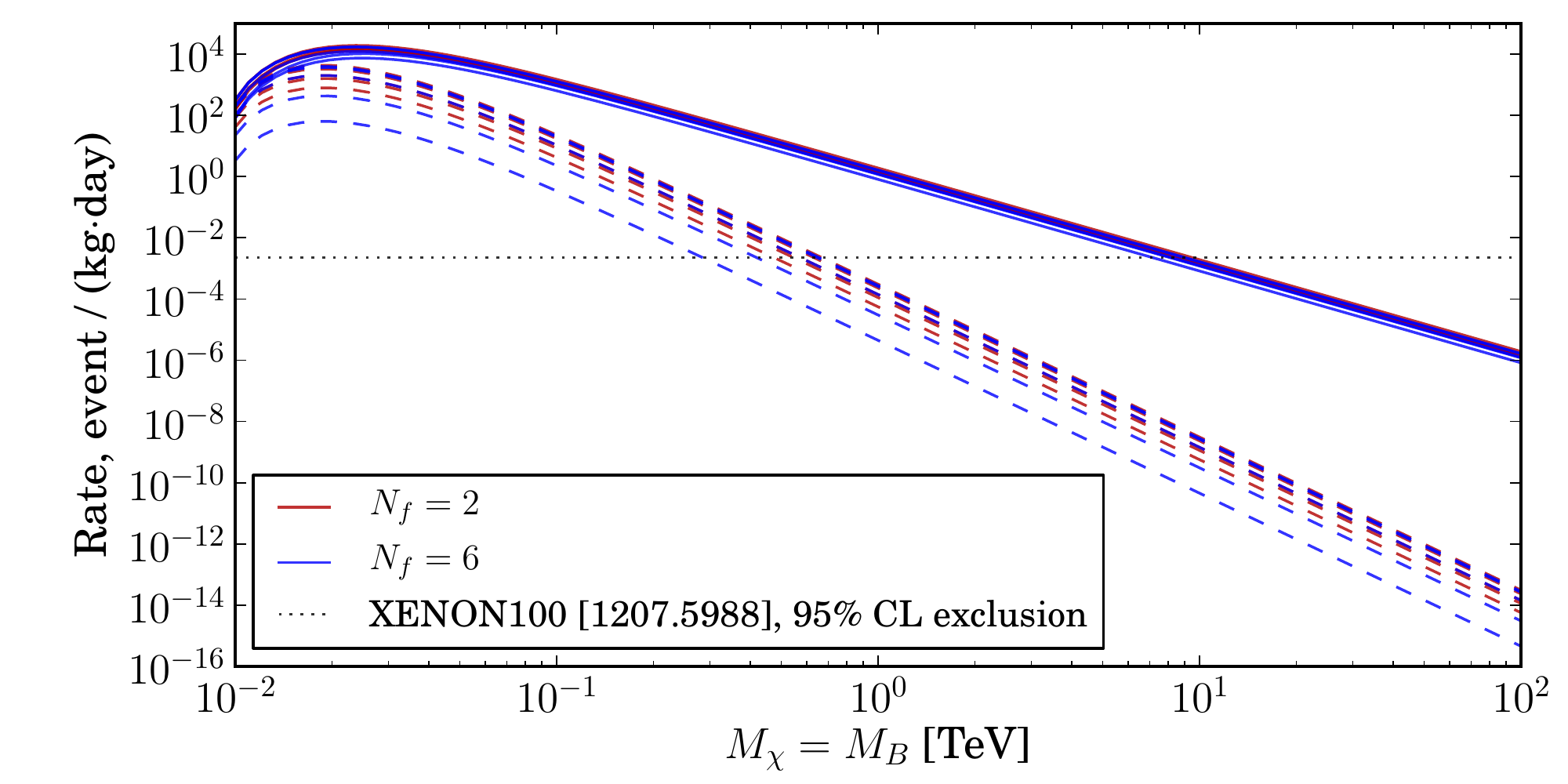}
  \caption{\label{fig:Exclusion}
    Calculated XENON100 event rates based on energy cuts and acceptance rates
    from Ref.~\cite{Aprile:2012nq} (solid lines).
    For comparison, we also show scattering rates using only the charge radius
    term, which is suppressed by two additional powers of $M_\chi$
    (dashed lines). The experimental upper bound on event rates, based on accumulated 2323.7
    kg$\cdot$days of exposure~\cite{Aprile:2012nq} are shown with the dotted lines.}
\end{figure}

We next compare our calculations of dark-matter parameters with the current experimental bounds on
the dark-matter-nucleus cross-sections in direct detection experiments.
Currently, the most stringent bound is provided by the XENON100 experiment~\cite{Aprile:2012nq},
in which hypothetical dark-matter particles are detected through their collisions with xenon nuclei
with $Z=54$ and $A=124\ldots136$, and which has accumulated 2323.7 kg$\cdot$days
of effective exposure. Two of the isotopes, ${}^{129}\text{Xe}$ and ${}^{131}\text{Xe}$,
have non-zero spin and are sensitive to the spin-dependent $M1$ interaction.
Their combined abundance constitutes approximately 1/2 in natural
xenon~\cite{handbook1986chemistry}.

In this section, we adopt a more conventional notation $M_\chi$ for the mass of the dark-matter
particle, and also denote its radius and magnetic moment with a subscript ``$\chi$''.
Figs. ~\ref{fig:Mag_Moment} and ~\ref{fig:Charge_Rad} show that
the anomalous moment and mean square charge radius vary little 
with the amount of the dark-matter mass
coming from the underlying fermion mass 
(and also vary little as $N_f$ is increased from $2$ to $6$).

The differential cross-section of a dark-matter fermion and a nucleus, to leading order in the
non-relativistic dark-matter velocity $v\ll 1$ is
\begin{equation}
\frac{d\sigma}{dE_R}
  = \frac{\overline{|\mathcal M_\text{SI}|^2} + \overline{|\mathcal M_\text{SD}|^2}}
         {16\pi (M_\chi + M_T)^2 E_R^\text{max}}\,,
\end{equation}
where $M_T$ is the mass of
the target nucleus, and $E_R^\text{max} = \frac{2M_\chi^2 M_T v^2}{(M_\chi + M_T)^2}$ is the maximal recoil energy
for given collision velocity $v$. The quantities and $\overline{|\mathcal M_\text{SI,SD}|}^2$ are
spin-(in)dependent amplitudes squared, averaged over initial and summed over final states:
\begin{align}
\label{eqn:amplSI_sq}
&\begin{aligned}
\overline{|\mathcal M_\text{SI}|^2}
  &= e^4 \, \big[Z F_c(Q)\big]^2 \, \Big(\frac{M_T}{M_\chi}\Big)^2\Big[
      \frac49 M_\chi^4  \langle r_{E\chi}^2\rangle^2 \\
&\quad\quad\quad  + \Big(1+\frac{M_\chi}{M_T}\Big)^2
        \kappa_\chi^2 \cot^2\frac{\theta_\text{CM}}2 \Big]\,,
\end{aligned}
\\
&\overline{|\mathcal M_\text{SD}|^2}
  = e^4 \, \frac23\Big(\frac{J+1}{J}\Big) \,
    \Big[\Big(A\frac{\mu_T}{\mu_n}\Big) \, F_s(Q)\Big]^2 \kappa_\chi^2\,.
\end{align}

Here, $Z$ and $A$ are the charge and atomic numbers of a specific xenon isotope,
$(\mu_T/\mu_n)$ is the nucleus magnetic moment expressed in Bohr magnetons
$\mu_n = \frac{e}{2m_n}$,
$F_{c,s}(Q)$ are its nuclear charge and spin form factors, respectively,
at the momentum transfer $Q\approx\sqrt{Q^2}=\sqrt{2M_T E_R}$,
and $\theta_\text{CM}$ is the scattering angle in the center-of-mass frame~\footnote{
  In comparison to the differential cross-section in Ref.~\cite{Banks:2010eh},
  Eq.~(\ref{eqn:amplSI_sq}) is the full spin-independent amplitude taking into account
  the non-trivial electric form factor of a DM particle,
  $G_{E\chi}(Q^2) = -\frac16 Q^2 r_{E\chi}^2 + {\mathcal O}(Q^4)$.
  With the Dirac radius set to zero,
  $r_{1\chi}^2 = r_{E\chi}^2 - \frac{3\kappa_\chi}{2m_\chi^2} = 0$,
  Eq.~(\ref{eqn:amplSI_sq}) reproduces identically the results in Ref.~\cite{Banks:2010eh}
  with $g_M=\frac12\kappa_\chi$ and zero electric dipole moment ($g_E=0$).}.
For non-relativistic velocities,
$\cot^2\frac{\theta_\text{CM}}2 = \Big(\frac{E_R^\text{max}}{E_R} - 1$\Big).

For the nuclear response form factors $F_{c,s}(Q^2)$, 
we use the following commonly accepted phenomenological
expressions ~\cite{Lewin:1995rx,Banks:2010eh}:
\begin{align}
|F_c(Q)|^2 &= 9\left|\frac{\sin(QR_c) - (QR_c)\cos(QR_c)}{(QR_c)^3}\right|^2 e^{-(QS)^2}\,,\\
|F_s(Q)|^2 &= \left\{\begin{array}{ll}
    0.047\,, &  2.55\le QR_s\le 4.5\,,\\
    \left|\frac{\sin(QR_s)}{QR_s}\right|^2\,, & \text{otherwise}\,,
  \end{array}\right.
\end{align}
where $R_c = 1.14 A^{1/3}\text{ fm}$, $R_s=1.00 A^{1/3}\text{ fm}$, and $S=0.9\text{ fm}$.
The nuclear response functions $F_{c,s}(Q^2)$ can also be evaluated using nuclear 
models, as was done in Ref.~\cite{Fitzpatrick:2012ix,Fitzpatrick:2012ib}.

Following Refs.~\cite{Banks:2010eh,Aprile:2012nq}, we compute the scattering rate for a range of
dark-matter particle masses with the recoil energies $E_R=6.6\ldots43.3\text{ keV}$:
\begin{equation}
\label{eqn:scatt_rate}
R = \frac{M_\text{detector}}{M_T}\frac{\rho_\text{DM}}{M_\chi}
  \int_{E_\text{min}}^{E_\text{max}}\,dE_R\,{\mathcal Acc}(E_R)\,
    \Big< v^\prime \, \frac{d\sigma}{dE_R}\Big>_f\,,
\end{equation}
where 
$\langle\cdot\rangle_f$ denotes averaging over the DM velocity
distribution~(\ref{eqn:dm_vel_distrib}),
$v^\prime = |\vec v - \vec v_\text{Earth}|$ is the dark-matter velocity with respect
to the detector (the Earth),
and ${\mathcal Acc}(E_R)$ is the recoil energy-dependent acceptance rate of the
detector~\cite{Aprile:2012nq}.
We assume the thermal distribution of velocities in the galactic dark-matter 
halo~\cite{Lewin:1995rx},
\begin{equation}
\label{eqn:dm_vel_distrib}
\frac {d^3n}{d\vec v^3} = f(\vec v) = \frac1{\pi^{3/2} v_0^3} e^{-\vec v^2 / v_0^2},
\quad \int_{|\vec v| < v_\text{esc}} d^3\vec v \;f(\vec v) \equiv 1\,,
\end{equation}
with $v_0=v_\text{Earth}=220\text{ km/s}$, $v_\text{esc}=544\text{ km/s}$, and
the dark-matter mass density $\rho_\text{DM}=0.3\text{ GeV/cm}^3$.
Finally, we average the expected scattering rate over the natural xenon isotopic abundances.

We show computed scattering rates in Fig.~\ref{fig:Exclusion} with solid lines. 
The accumulated XENON100 statistics~\cite{Aprile:2012nq} exclude composite dark matter
particles with $M_\chi\lesssim10\text{ TeV}$.
With fixed values of the dimensionless quantities $M_\chi^2 \langle r^2 \rangle$ and $\kappa$
computed on a lattice, the differential cross-section scales as
\begin{equation}
\frac{d\sigma}{dE_R}
  \sim {\mathcal A} \frac{(M_\chi^2 \langle r_{E\chi}^2 \rangle )^2}{M_\chi^{4}}
  + {\mathcal B}\frac{\kappa_\chi^2}{M_\chi^2}\,.
\end{equation}
The charge radius contribution is suppressed as $M_\chi^{-2}$ relative to
the magnetic moment contribution and becomes negligible with growing $M_\chi$.
In the scattering rate shown in Fig.~\ref{fig:Exclusion}, both contributions are additionally 
suppressed by the DM particle number density $\rho_\text{DM}/M_\chi$ as $M_\chi\to\infty$ 
(see Eq.(\ref{eqn:scatt_rate})).
The large-$M_\chi$ scaling of the charge radius term is shown in Fig.~\ref{fig:Exclusion} 
with the dashed lines; it is evident that the total scattering rate (solid lines) is
dominated by the magnetic moment term for dark matter masses $M_\chi\gtrsim25\text{ GeV}$. 
Even if one were to make the charge radius as much as an order of magnitude larger 
by reducing the PNGB mass (see the discussion following Fig.~\ref{fig:Charge_Rad}), 
its contribution would still be negligible at $M_B = 10\text{ TeV}$, 
the lower limit of the allowed region.

\paragraph{\textbf{Discussion}}
We have studied the electromagnetic form factors of electroweak-neutral dark-matter baryons in
an $SU(3)$ gauge theory with $N_f = 2$ and $6$ $SU(2)_L$-singlet fermions, with charge
assignments $+2/3$ and $-1/3$ (one pair or three pairs). These baryons have the desired
properties of dark matter since they are stable, electroweak neutral, and can explain the relic
density through the same early universe sphaleron process that describes baryogenesis.

Of particular interest to direct detection experiments are the anomalous magnetic moment and
mean square charge radius of the dark-matter baryon.  These parameters determine the observed
cross-section with nuclei (in this work, we primarily focus on xenon) due to (dominant)
single-photon exchange. The contribution from the dark-matter anomalous moment dominates the
charge-radius contribution for $M_\chi\gtrsim25 \text{GeV}$. 
However, in our calculation the charge radius $\langle r_E^2\rangle$ turns out to be
particularly small, much smaller than it is in QCD.
Exploring smaller quark mass regions may change this balance and make the charge radius 
more relevant for the direct detection of the dark matter.

Examining the dark matter exclusion plots in light 
of the latest XENON100 results~\cite{Aprile:2012nq}, we conclude that in these
theories, dark-matter masses less than $10 \ \text{TeV}$ are excluded.
We have so far seen little dependence on $N_f$. 
It will be interesting to see whether this begins to change,
even continuing to neglect disconnected quark contractions, 
as $N_f$ is increased toward the edge of the conformal window 
($N_{f} \approx 10-12$ for an $SU(3)$ gauge theory with fermions 
in the fundamental representation).
When the disconnected contractions are included, additional $N_f$ dependence
will arise simply from the counting of these loop contributions.

As we have shown in this work, with non-zero magnetic moment the
experimental constraints on the dark matter mass are quite stringent.
This naturally motivates the consideration of even $N_c$ theories, in
which the baryons are bosonic and thus have no magnetic moment.
Interactions can be further suppressed if the charge assignments are
symmetric in such a way that the charge radius vanishes (see e.g.
\cite{Kribs:2009fy}), making the electromagnetic polarizabilities the
dominant interactions.  Some initial lattice work on the zero-temperature dynamics of such theories has been carried out in \cite{Lewis:2011zb,Hietanen:2012sz}, and we are currently planning similar calculations with an eye towards dark matter form factors.

\paragraph{\textbf{Acknowledgements}}
We thank the LLNL Multiprogrammatic and Institutional Computing program for Grand Challenge
allocations and time on the LLNL BlueGene/L (uBGL) supercomputer as well as on the LLNL Hera,
Atlas, and Zeus computing clusters. We thank LLNL for funding from LDRD~10-ERD-033 and 
LDRD~13-ERD-023.
The LSD collaboration would like to thank Graham Kribs for his valuable input 
on this effort and for providing insightful comments on the manuscript draft.
MIB would like to thank Graham Kribs for many illuminating and insightful discussion 
throughout this work, along with the hospitality of the University of Oregon particle theory group.
SNS would like to thank Wick Haxton for helpful discussions on nuclear response form factors.
Several of the authors would also like to thank the Kavli Institute for Theoretical Physics 
and the organizers of the program 
``Novel Numerical Methods for Strongly Coupled Quantum Field Theory and Quantum Gravity'', 
where much of this work was developed.
This work has been supported by the U.~S.~Department of Energy under Grant Nos.~DE-FG02-04ER41290 
(D.S.), DE-FG02-91ER40676 (R.C.B., M.C., C.R.), DE-FG02-92ER-40704 (T.A.) 
and Contracts DE-AC52-07NA27344 (LLNL), DE-AC02-06CH11357 (Argonne Leadership Computing Facility), 
and DE-AC02-07CH11359 (Fermi Research Alliance, LLC), and by the National Science Foundation 
under Grant Nos.~NSF PHY11-00905 (G.F., M.L., G.V.) and PHY11-25915 
(Kavli Institute for Theoretical Physics).
S.N.S was supported by the Office of Nuclear Physics in the US Department of Energy's
Office of Science under Contract DE-AC02-05CH11231.

\bibliography{LSD-FF}

\end{document}